\documentclass[twocolumn,aps,prd,10pt,superscriptaddress,longbibliography,floatfix,citeautoscript]{revtex4-2}
\usepackage{graphicx}
\usepackage{amsmath}
\usepackage{threeparttable}
\usepackage{amssymb}
\usepackage{booktabs}
\usepackage{bm}
\usepackage{dcolumn}
\usepackage{glossaries}
\usepackage{graphicx}
\usepackage{multirow}
\usepackage{fancyvrb}
\usepackage[linesnumbered,ruled]{algorithm2e}
\usepackage{siunitx}
\usepackage{soul}
\usepackage{textcomp}
\usepackage{glossaries}
\makeglossaries
\usepackage[usenames,dvipsnames,svgnames]{xcolor}
\usepackage{hyperref}
\usepackage{float}
\usepackage{array}
\usepackage{makecell}
\hypersetup{
    pdfnewwindow=true,      
    colorlinks=true,        
    linkcolor=Blue,         
    citecolor=Blue,         
    filecolor=Blue,         
    urlcolor=Blue           
}

\usepackage{listings}	
\newacronym{md}{MD}{molecular dynamics}
\newacronym{npt}{NPT}{isothermal-isobaric ensemble}
\newacronym{vasp}{VASP}{Vienna Ab initio Simulation Package}
\newacronym{dft}{DFT}{density functional theory}
\newacronym{paw}{PAW}{Projected augmented wave}
\newacronym{gga-pbe}{GGA-PBE}{Generalized Gradient Approximation Perdew-Burke-Ernzerhof}
\newacronym{gpumd}{GPUMD}{Graphics Processing Units Molecular Dynamics}
\newacronym{nep}{NEP}{neuroevolution potential}
\newacronym{snes}{SNES}{separable natural evolution strategy}
\newacronym{rmse}{RMSE}{root mean square error}
\newacronym{nve}{NVE}{micro-canonical}
\newacronym{nvt}{NVT}{canonical}

\newacronym{tan}{$ \theta$-TaN}{$ \theta$-TaN}

\newacronym{ace}{ACE}{atomic cluster expansion}

\newacronym{cmd}{CMD}{classical molecular dynamics}
\newacronym{pimd}{PIMD}{path integral molecular dynamics}
\newacronym{mlp}{MLP}{Machine-learned potential}
\newacronym{mae}{MAE}{mean absolute error}
\newacronym{vdw}{vdW}{van-der-Waals}
\newacronym{nn}{NN}{neural network}
\newacronym{rdf}{RDF}{Radial Distribution Function}

\DeclareSIUnit\angstrom{\text{Å}}
\DeclareSIUnit{\atom}{atom}
\DeclareSIUnit{\step}{step}
\DeclareSIUnit{\atomstepsecond}{\atom\step\per\second}

\newcolumntype{d}{D{.}{.}{-1}}

\begin{document}

\title{Anisotropic Tensile Strength and Fracture Mechanism of $\theta$-TaN: A Machine-Learning Potential Molecular Dynamics Study}

\author{Chenyang Cao}
\affiliation{Department of Physics, University of Science and Technology Beijing, Beijing 100083, China}

\author{Hongfei Li}
\affiliation{ Beijing Computing Center Co. Ltd, Beijing Academy of Science and Technology, Beijing 100094,China}

\author{Shuo Cao}
\email{icaoshuo@outlook.com}
\affiliation{College of Physical Science and Technology, Bohai University, Jinzhou 121013, China}

\date{\today}
\begin{abstract}
$\theta$-phase tantalum nitride ($\theta$-TaN) combines metallic conductivity with exceptionally high thermal conductivity, making it a potential material for device thermal management and interconnect applications. However, its tensile strength and fracture behavior remain unclear. Here, we investigate the anisotropic tensile response and fracture mechanism of $\theta$-TaN using neuroevolution-potential molecular dynamics simulations. Size-convergence tests show that a $20~\mathrm{nm}$ long model is sufficient for reliable prediction, and the mechanical parameters vary by less than 3.5\% over the strain-rate range of $10^7$–$10^9~\mathrm{s^{-1}}$. The results reveal strong tensile anisotropy. The $c$-axis direction ($[0001]$) shows a higher strength of $80.10~\mathrm{GPa}$ and modulus of $748.63~\mathrm{GPa}$, but a lower fracture strain of 15.02\%. In contrast, the $a$-axis direction ($[2\bar{1}\bar{1}0]$) shows a lower strength of $56.87~\mathrm{GPa}$ and modulus of $570.74~\mathrm{GPa}$, but a higher fracture strain of 17.71\%. From 300 to 900 K, the mechanical properties decrease nearly linearly, while more than 73\% of the 300 K strength is retained at 900 K. Fracture occurs without observable dislocation activity and is governed by cleavage-plane selection: ${10\bar{1}0}$ prismatic planes under $a$-axis tension and the $(0001)$ basal plane under $c$-axis tension. Atomic displacement analysis shows that local separation and microvoid formation precede macroscopic crack growth, indicating a brittle fracture process driven by local bond-network instability. These results provide atomic-scale mechanical data for assessing the reliability of $\theta$-TaN in thermal management applications.

\end{abstract}

\maketitle

\section{Introduction}

As the power density of electronic devices continues to increase, thermal management \cite{Li2025amr,Mengesha2025DM,cui2020jmcc} has become an important issue in device design and long-term reliability. High thermal conductivity substrates, including diamond\cite{diamond2023MTP}, cubic boron arsenide (BAs)\cite{BAs2018science,BAs2025matter,BAs2026MNE}, and boron nitride (BN)\cite{BN2019SA,BN2019S,BN2024v}, have been widely considered as possible alternatives to conventional packaging-based thermal management\cite{review2025MR} strategies. However, their integration with established semiconductor processing routes remains challenging, which has limited their practical use. In contrast to these insulating or semiconducting materials, $\theta$-phase tantalum nitride ($\theta$-TaN) is a metallic nitride\cite{Li2026science}. Recent reports by Kundu \textit{et al}.\cite{Kundu2021PRL,Kundu2024PRL} and Li \textit{et al}.\cite{Li2026science} showed that its room-temperature thermal conductivity can reach approximately $1000~\mathrm{W \cdot m^{-1} \cdot K^{-1}}$. Another notable feature of $\theta$-TaN is its WC-like crystal structure. Since tungsten carbide (WC)\cite{WC2025ACTA,WC2024jecs,WC2025ci} is a representative hardmetal with high strength and hardness, this structural similarity raises the possibility that $\theta$-TaN may also exhibit useful mechanical properties. Therefore, $\theta$-TaN offers an unusual combination of metallic character, ultrahigh thermal conductivity, and possible mechanical robustness, making it relevant for next-generation thermal management and interconnect applications.

However, the use of $\theta$-TaN in practical devices cannot be judged from its thermal transport properties alone. Thin films are also exposed to thermal stress, mechanical loading, and repeated cycling during service. These conditions are directly linked to the mechanical response of the material. At present, there is still little information on the tensile strength, fracture behavior, and failure mode of $\theta$-TaN. This makes it difficult to assess whether its high thermal conductivity can be matched with sufficient reliability for structural or functional use.

Studying fracture behavior under large deformation and over extended length and time scales remains challenging for first-principles density functional theory molecular dynamics simulations, because AIMD is usually limited by small system sizes and short simulation times. Conventional empirical potentials can reach the required length and time scales, but they often lack the accuracy needed to describe complex events such as bond breaking. Machine learning potentials (MLPs) provide a practical way to address this problem. They allow large-scale molecular dynamics (MD) simulations with an efficiency close to that of empirical potentials, while retaining an accuracy that approaches first-principles calculations.

\begin{figure*}[htbp]
\centering
\includegraphics[width=1.15\columnwidth]{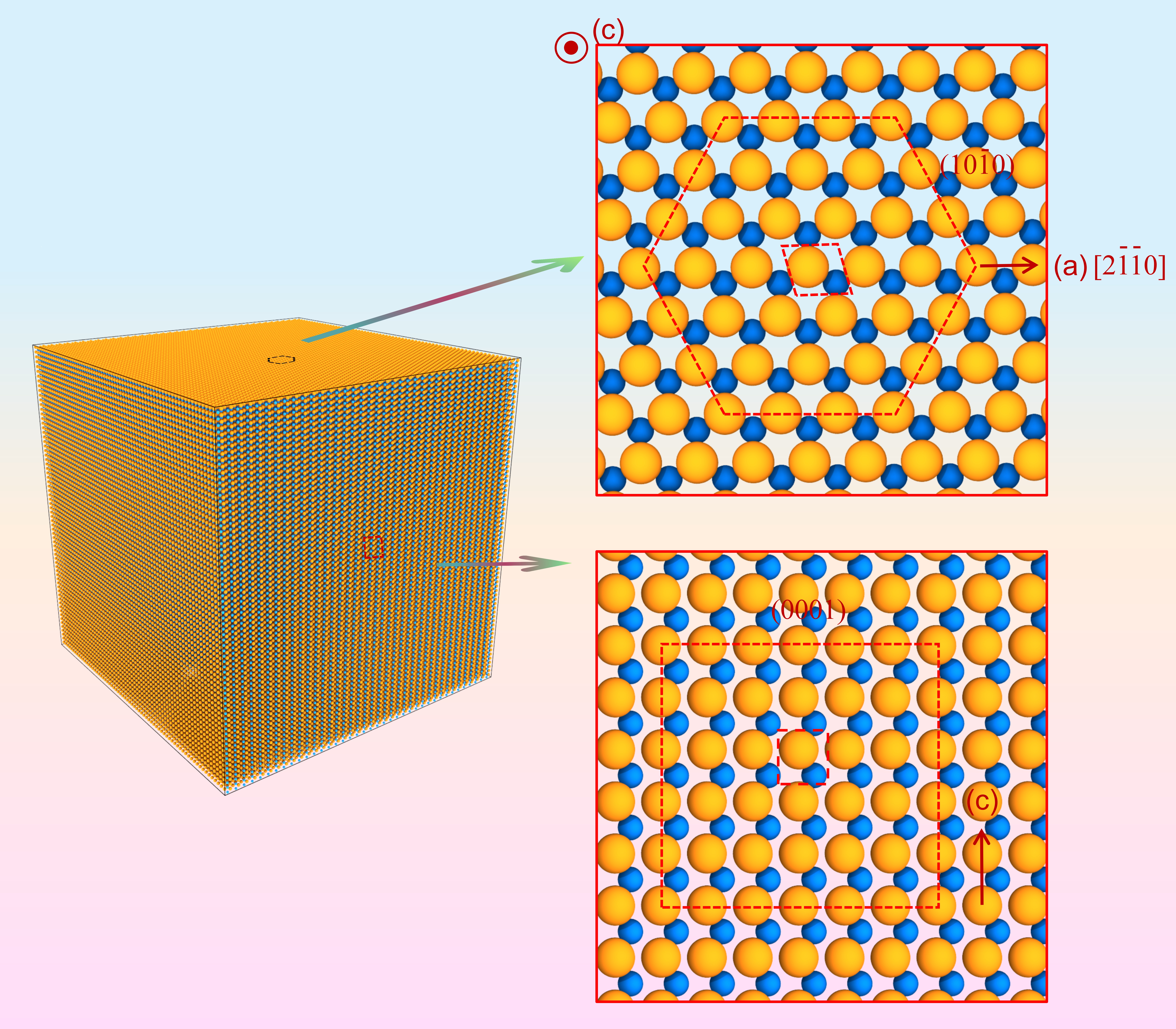}
\caption{\textbf{Atomic model and crystallographic orientations of $\theta$-TaN.} The left part shows the complete supercell used in the tensile simulations, with a size of approximately $20~\mathrm{nm} \times 20~\mathrm{nm} \times 20~\mathrm{nm}$ and 731,952 atoms. The upper-right view is along the $c$ axis and shows the atomic arrangement in the basal plane, where the $a$-axis loading direction $[2\bar{1}\bar{1}0]$ and the $(10\bar{1}0)$ prismatic plane are marked. The lower-right view is along the $a$ axis and shows the $(0001)$ basal plane and the $c$-axis loading direction $[0001]$. The red dashed boxes indicate the orientation of the hexagonal primitive cell within the supercell.}
\label{fig:atom model}
\end{figure*}

Among commonly used MLPs, including Deep Potential (DP)\cite{wang2018deepmd}, Gaussian approximation potential (GAP)\cite{bartok2010gaussian}, moment tensor potential (MTP)\cite{LI2022MTP}, recursive embedded-atom neural networks (REANN)\cite{REANN2021PRL}, neural equivariant interatomic potentials (NequIP)\cite{nequip2026DD,nequip2022nc}, and message-passing atomic cluster expansion (MACE)\cite{batatia2022mace}, the neuroevolution potential (NEP)\cite{song2024general,Fan2019prb,Fan2021prb,Fan2017cpc}, provides a useful balance between computational efficiency and accuracy. NEP has been applied to the tensile deformation of various materials\cite{LIU2026MSEA,REN2026ACTA,Wang2025jap,Shi2024JPCM,YU2024jmps,YING2023EML,QIN2026111741}. This feature is particularly important for the present study, because it allows MD simulations to cover sufficiently large system sizes\cite{song2024general} and to examine the size convergence of the mechanical properties\cite{YING2023EML}. Based on machine learning potential methods\cite{Lin2024npj}, previous studies on the tensile fracture behavior of nitrides and borides have established a relatively complete analysis framework, including tests of size convergence, strain-rate dependence, and temperature dependence. The present work follows this framework.

In this work, we employ the NEP model together with large-scale molecular dynamics simulations to investigate the tensile mechanical properties of $\theta$-TaN. The aim is to provide missing mechanical data for this material, including tensile strength, tensile modulus, and fracture strain. We first perform a size-convergence analysis to determine the simulation cell size required for reliable prediction of the mechanical properties. We then examine the strain-rate and temperature dependences, and establish the relationships between strength, modulus, and temperature for $\theta$-TaN. Under representative loading conditions, frame-by-frame structural analysis and atomic displacement mapping are further carried out to identify the brittle fracture characteristics, fracture planes, and crack evolution process. Based on these analyses, we discuss the atomic-scale relationship among local displacement separation, damage accumulation, and bond breaking before macroscopic failure. Finally, the tensile strength of $\theta$-TaN is compared with those of several representative structural materials to clarify its relative mechanical position among high-performance ceramics and transition-metal compounds. This work establishes a systematic atomic-scale mechanical reliability framework for $\theta$-TaN, covering both macroscopic tensile properties and microscopic fracture evolution.

\section{Models and Methods}

\subsection{Atomistic models}

The crystal structure of $\theta$-TaN belongs to the hexagonal crystal system, with the space group $P\bar{6}m2$. It has the same structure type as tungsten carbide (WC). The Ta atoms form a simple hexagonal sublattice, while the N atoms occupy half of the trigonal-prismatic interstitial sites. The Ta and N atomic layers are alternately stacked along the $c$ axis. In this work, the initial unit cell was constructed using the lattice constants obtained from self-consistent density functional theory (DFT) calculations, with $a = 2.926~\text{\AA}$ and $c = 2.873~\text{\AA}$.

To investigate the anisotropic tensile response of $\theta$-TaN, two representative crystallographic directions were selected as the tensile loading directions: the $a$-axis direction, $[2\bar{1}\bar{1}0]$, which lies in the basal plane and points from the center of the hexagon to one of its vertices, and the $c$-axis direction, $[0001]$. These two directions represent typical in-plane and out-of-plane directions in the hexagonal structure, respectively. This choice also allows direct comparison with the reported anisotropy in thermal conductivity\cite{Kundu2024PRL} for this material.

Fig.~\ref{fig:atom model} shows the atomic models constructed in this work. The left panel shows the complete supercell, with dimensions of approximately $20~\mathrm{nm} \times 20~\mathrm{nm} \times 20~\mathrm{nm}$ and a total of 731,952 atoms. The choice of this cell size is justified by the size-convergence analysis in Section ~\ref{size_sec}. The upper-right panel shows the structure viewed along the $c$ axis, highlighting the atomic arrangement of the hexagonal unit cell in the basal plane. The labeled $(10\bar{1}0)$ plane corresponds to the side face of the hexagonal prism, namely the $m$ prismatic plane, and lies along the edge direction of the hexagon. The $a$-axis direction, $[2\bar{1}\bar{1}0]$, points from the center of the hexagon to one of its vertices and is used as the in-plane tensile loading direction in this study. The lower-right panel shows the structure viewed along the $a$ axis, displaying the $(0001)$ basal plane perpendicular to this direction. The $c$-axis direction, corresponding to the out-of-plane tensile loading direction used in this work, is also marked. The red dashed boxes indicate the orientation relationship between the hexagonal primitive cell and the supercell.

\subsection{Neuroevolution potential model}
\subsubsection{The NEP Formalism}

In this work, the potential energy surface of $\theta$-TaN was constructed using the neuroevolution potential (NEP) framework \cite{song2024general,Fan2017cpc,Fan2019prb,Fan2021prb}. This method combines a single-hidden-layer feedforward neural network with the separable natural evolution strategy (SNES) optimization algorithm\cite{Schaul2011}. The model parameters are iteratively optimized during training, allowing the interatomic interactions to be described with high accuracy.

In the NEP framework, the total energy of a system is expressed as the sum of the site energies $U_i$ of all atoms. For a central atom $i$, the site energy is predicted by a neural network as

\begin{equation}
\label{equation:u_i}
U_{i} = \sum_{\mu=1}^{N_{\rm neu}} \omega_\mu^{(1)} \tanh\left(\sum_{\nu=1}^{N_{\rm des}}\omega_{\mu\nu}^{(0)}q_{\nu}^i-b_{\mu}^{(0)}\right)-b^{(1)},
\end{equation}

where $q_{\nu}^{i}$ is the $\nu$-th component of the descriptor vector of atom $i$. The hyperbolic tangent function, $\tanh$, is used as the activation function in the hidden layer to introduce nonlinear mapping capability. $N_{\mathrm{des}}$ and $N_{\mathrm{neu}}$ denote the descriptor dimension and the number of hidden-layer neurons, respectively. The network parameters, ${\omega, b}$, are iteratively optimized using the SNES algorithm, so that the network output progressively approaches the reference values obtained from first-principles calculations.

To characterize the local geometric environment around each atom, the descriptor vector $\mathbf{q}^{i}$ in NEP contains both radial and angular components. Its construction follows the basic ideas of the Behler–Parrinello symmetry functions \cite{behler2007prl} and the smooth overlap of atomic positions (SOAP) method~\cite{bartok_2013,2019_Miguel_soap}. Specifically, the radial component $q_{n}^{i}$, which describes two-body interactions, is obtained by summing a set of radial basis functions $g_{n}(r_{ij})$:
\begin{equation}
\label{equation:rad_des}
q_{n}^i = \sum_{j\neq{i}}g_n(r_{ij}),~0 \leq {n} \leq n_{\rm max}^{\rm R},
\end{equation}

The two-body radial information alone is often insufficient to fully describe the complex local bonding environment in $\theta$-TaN. Therefore, NEP further introduces angular descriptors to account for three-body and higher-order correlations. The three-body component $q_{nl}^{i}$ ($0 \le n \le n_{\rm max}^{\rm A}$, $1 \le l \le l_{\rm max}$) is defined as:
\begin{equation}
\label{equation:ang_des}
q_{nl}^i =\frac{2l+1}{4\pi}\sum_{j\neq{i}}\sum_{k\neq{i}}g_n(r_{ij})g_n(r_{ik})P_l(\cos\theta_{ijk}),
\end{equation}

Here, $\theta_{ijk}$ is the angle centered at atom $i$ and formed by the two bonds $ij$ and $ik$, while $P_l$ is the Legendre polynomial of order $l$. It should also be noted that the radial basis functions used in the angular components can adopt a cutoff radius $r_{\rm c}^{\rm A}$ different from that used for the two-body terms. This allows a balance between computational efficiency and descriptor accuracy in practical calculations.

\subsubsection{Datasets for training the NEP model}

To accurately evaluate the mechanical response of $\theta$-TaN under tensile loading, a training dataset specifically designed for tensile simulations was constructed. For the perfect crystal, configurations were sampled from NPT, NVT, and selected NVE ensembles over the temperature range of 300–900 K. To improve the ability of the potential to describe local structural perturbations and deformed configurations, lattice strains from $-6\%$ to $+6\%$ with an interval of $1\%$, atomic displacement perturbations, and shear strains were applied to the perfect structures.

Since the main objective of this work is to study the anisotropic tensile behavior of $\theta$-TaN along the $a$ and $c$ axes, the training dataset further included uniaxially strained configurations along the $a$ axis ($[2\bar{1}\bar{1}0]$) and the $c$ axis ($[0001]$) at different temperatures from 300 to 900 K. The strain range was set to $1\%$–$25\%$. These configurations were used to represent the potential energy surface along the two representative crystallographic directions, covering the deformation process from the elastic regime to the large-strain region close to fracture, together with its temperature dependence.

The primitive cell of $\theta$-TaN has hexagonal symmetry. A hexagonal-to-tetragonal cell transformation was first performed, increasing the number of atoms in the cell from 2 to 4. This transformed tetragonal cell was then used to construct tensile supercells along the $a$- and $c$-axis directions. All subsequent supercells were built based on this tetragonal cell. For temperature sampling and shear-strain configurations, the training data were generated using 128-atom supercells. In addition, generalized stacking-fault configurations were constructed along relevant slip systems for several surface orientations, including the basal, $m$-prismatic, and $a$-prismatic planes. These configurations also contained 128 atoms.

To further improve the accuracy of the model in the large-deformation regime, an active-learning strategy was used to iteratively enrich the dataset. This procedure focused on configurations subjected to $1\%$–$25\%$ uniaxial tensile strain along the $a$ and $c$ axes at temperatures from 300 to 900 K. Specifically, molecular dynamics tensile simulations were first carried out at each temperature and along each loading direction using the initially trained NEP model. Strained configurations with high prediction uncertainty, or with relatively large deviations from the DFT reference values, were then identified and added back to the training dataset for retraining. This process was repeated until the predicted energies and forces became converged over the $1\%$–$25\%$ strain range for both tensile directions and all considered temperatures.

The final training dataset contained 1,053 configuration snapshots and 115,752 atoms, providing a diverse set of samples for model training. Among them, 646 configurations, accounting for 61.3\% of the dataset, were thermal-vibration configurations without external strain in the temperature range of 300–900 K. The remaining 407 configurations, accounting for 38.7\%, consisted of lattice-perturbation configurations, shear-strain configurations, uniaxially strained configurations along the $a$ and $c$ axes over the $1\%$–$25\%$ strain range, including those added through active learning, and generalized stacking-fault configurations.

To independently evaluate the generalization ability of the model, an additional test dataset was constructed. It contained 400 compressed configurations sampled from NPT ensembles at hydrostatic pressures of 1, 2, 5, and 10 GPa over the temperature range of 300–900 K. The loading mode in the test dataset was hydrostatic compression, which differs from the main strain paths in the training dataset, namely uniaxial tension and shear. Therefore, this test dataset can be used to assess the predictive capability of the NEP model for configurations outside the main training domain.

The reference dataset was obtained from first-principles calculations using the \textsc{VASP} package\cite{vasp}, including total energies, atomic forces, and virial stress tensors. The projector augmented-wave (PAW) method \cite{paw1,paw2} was used together with the GGA-PBE exchange-correlation functional\cite{GGA-PBE}. The plane-wave cutoff energy was set to 600 eV, and the $\text{Ta}\_{\text{pv}}$ pseudopotential was used for Ta. The Brillouin zone was sampled using $\Gamma$-centered $k$-point meshes with a maximum spacing of $0.2~\text{\AA}^{-1}$. Electronic self-consistent calculations were performed using the Normal algorithm, with an energy convergence criterion of $10^{-6}~\mathrm{eV}$, a maximum of 80 electronic steps, and a Gaussian smearing width of $0.02~\mathrm{eV}$.

\subsubsection{Trained NEP model}

The NEP model was trained using the \textsc{GPUMD} package \cite{fan2022gpumd}. The hyperparameters used in the training are summarized in Table~\ref{table:hyper}. The model parameters were optimized using the SNES algorithm. The loss function was defined as a weighted sum of the root-mean-square errors (RMSEs) of energy, force, and virial, with the weights $\lambda_e$, $\lambda_f$, and $\lambda_v$ set to 1.0, 1.0, and 0.1, respectively.

Based on convergence tests, both the radial cutoff radius $r_c^R$ and the angular cutoff radius $r_c^A$ were set to $5~\text{\AA}$ to minimize the fitting errors. The numbers of radial and angular basis functions, $N_{\rm bas}^R$ and $N_{\rm bas}^A$, were both set to 8, and the corresponding expansion orders, $n_{\max}^R$ and $n_{\max}^A$, were also set to 8. To describe the many-body interactions in $\theta$-TaN, the expansion orders of the three-body and four-body angular descriptors, $l_{\max}^{3b}$ and $l_{\max}^{4b}$, were set to 4 and 2, respectively. The hidden layer contained 50 neurons, i.e., $N_{\rm neu}=50$. During SNES optimization, the population size $N_{\rm pop}$ was set to 50, and the total number of generations $N_{\rm gen}$ was set to $3\times10^5$.

\begin{table}[htb]
\centering
\setlength{\tabcolsep}{2.5mm}
\caption{\textbf{Hyperparameters used for training the NEP model of $\theta$-TaN.} The table summarizes the cutoff radii, descriptor settings, neural-network size, optimization parameters, and loss-function weights.}
\label{table:hyper}
\begin{tabular}{lllllllll}
\hline
\hline
Parameter & Value & Parameter & Value\\
\hline
$r_{\rm c}^{\rm R}$ & 5 \AA & $r_{\rm c}^{\rm A}$ & 5 \AA \\
$n_{\rm max}^{\rm R}$ & 8 & $n_{\rm max}^{\rm A}$ & 8 \\
$N_{\rm bas}^{\rm R}$ & 8 & $N_{\rm bas}^{\rm A}$ & 8 \\
$l_{\rm max}^{\rm 3b}$ & 4 & $l_{\rm max}^{\rm 4b}$ & 2 \\
$N_{\rm neu}$ & 50  & $N_{\rm gen}$ & $3 \times 10^5$ \\
$\lambda_{\rm e}$ & 1.0  & $N_{\rm pop}$ & 50 \\
$\lambda_{\rm f}$ & 1.0 & $\lambda_{\rm v}$ & 0.1 \\
\hline
\hline
\end{tabular}
\end{table}

\subsection{The MLP-based tensile simulations}

Molecular dynamics (MD) tensile simulations were then carried out with the trained NEP potential to evaluate the mechanical response of $\theta$-TaN. All MD simulations were performed using GPUMD. The time step was $2~\mathrm{fs}$, and periodic boundary conditions were applied in the three cell directions.

\begin{figure*}
\centering
\includegraphics[width=1.6\columnwidth]{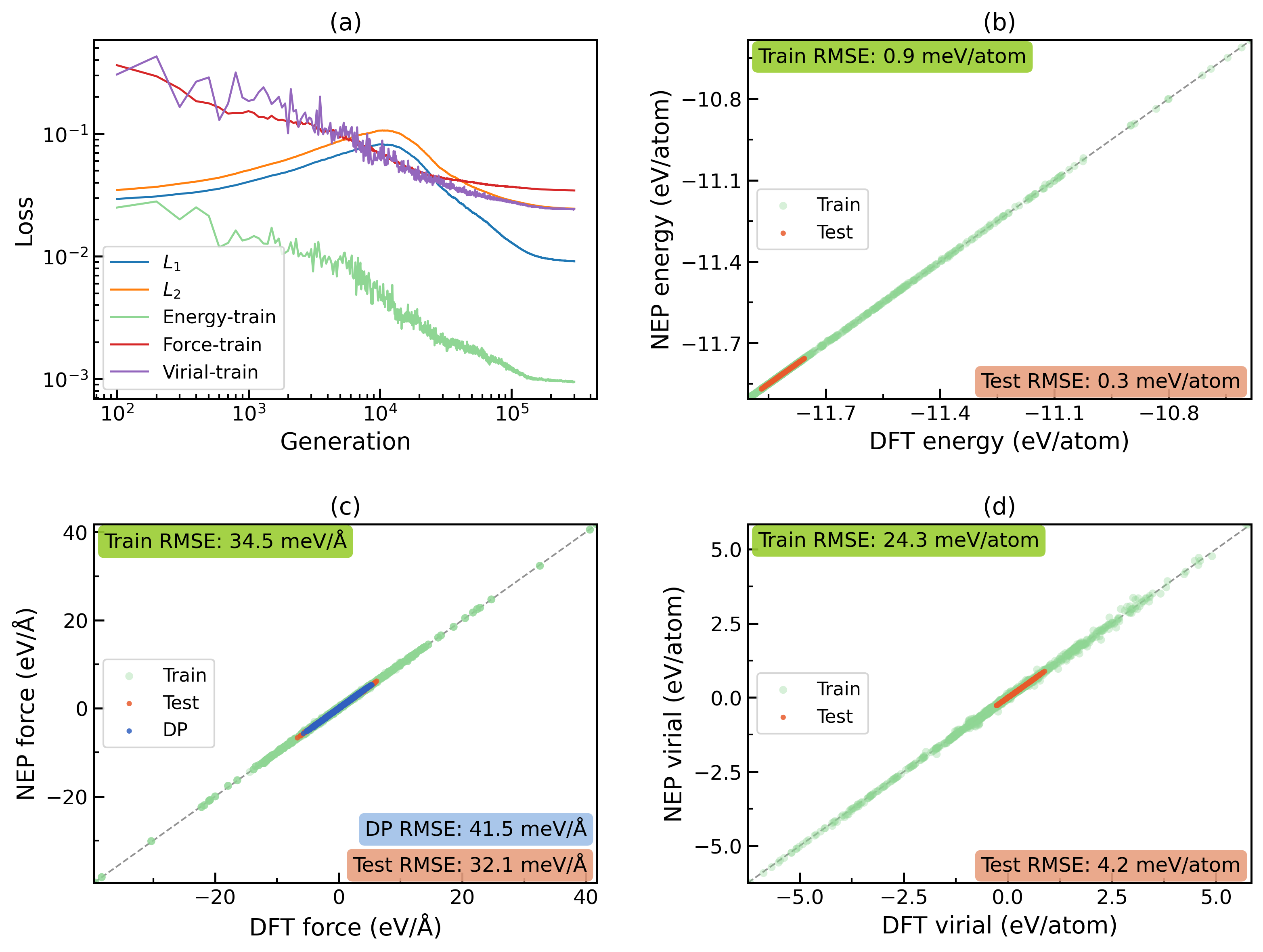}
\caption{\textbf{Accuracy validation of the trained NEP model for $\theta$-TaN.} (a) Evolution of the training loss during SNES optimization. (b) Comparison between NEP predictions and DFT reference values for energy on the training and test datasets. (c) Comparison of atomic forces, including the training dataset, test dataset, and external DP dataset\cite{ZONG2026ijhmt}. (d) Comparison between NEP predictions and DFT reference values for virial on the training and test datasets. The corresponding RMSE values are shown in each panel. }
\label{fig:rmse}
\end{figure*}

Each simulation contained an equilibration step and a tensile-loading step. The supercell was first equilibrated at the target temperature for $1~\mathrm{ns}$ in the NPT ensemble. The Bussi–Donadio–Parrinello thermostat and stochastic cell rescaling barostat were used during this step. The target pressures along the three cell directions were set to zero, so that the initial structure could relax to a nearly stress-free state. After equilibration, tensile strain was applied along either the $a$ axis ($[2\bar{1}\bar{1}0]$) or the $c$ axis ($[0001]$) at a prescribed strain rate. During loading, the temperature was kept at the target value by the same thermostat. The barostat was used to relax the transverse stresses, giving a quasi-uniaxial tensile condition.

To examine possible size effects, five supercell sizes were considered: 5, 10, 20, 30, and $40~\mathrm{nm}$. Strain-rate effects were tested using $1\times10^7$, $5\times10^7$, $1\times10^8$, $5\times10^8$, and $1\times10^9~\mathrm{s^{-1}}$. The temperature-dependent tensile response was calculated from 300 to $900~\mathrm{K}$ at an interval of $100~\mathrm{K}$. The elastic modulus, tensile strength, and fracture strain were extracted from the resulting stress–strain curves. The fracture strain was defined as the strain corresponding to the maximum
tensile stress.

The initial crystal structures, orientation transformations, and supercells were prepared using VASPKIT\cite{VASPKIT} and Atomsk\cite{Atomsk}. Atomic configurations from the MD trajectories were visualized and analyzed using OVITO\cite{ovito}. NepTrainKit\cite{NEPTRAINKIT} was used to evaluate the NEP fitting errors for energy, force, and virial, and to select configurations during active learning.

\section{Results and discussion}

\subsection{Benchmark}

To assess the reliability and transferability of the trained NEP model, we validated it from three aspects: static fitting accuracy, tensile response, and the energetic prediction of key metastable configurations.

We first examined the fitting accuracy of the NEP model on the training and test datasets. As shown in Fig.~\ref{fig:rmse}(a), the loss function decreases continuously with increasing training generations and then approaches convergence. This indicates that the SNES optimization is stable and that the model parameters are well converged. Figs.~\ref{fig:rmse}(b)--(d) compare the NEP predictions with the DFT reference values for energy, force, and virial in both datasets. For the training dataset, the root-mean-square errors (RMSEs) of energy, force, and virial are $0.9~\mathrm{meV/atom}$, $34.5~\mathrm{meV/\text{\AA}}$, and $24.3~\mathrm{meV/atom}$, respectively. For the independent test dataset, the corresponding RMSEs are $0.3~\mathrm{meV/atom}$, $32.1~\mathrm{meV/\text{\AA}}$, and $4.2~\mathrm{meV/atom}$, respectively. This test dataset contains 400 configurations sampled from NPT simulations at hydrostatic pressures of 1, 2, 5, and $10~\mathrm{GPa}$ over the temperature range of 300–900 K. The test errors are comparable to those of the training dataset, and some of them are even lower, indicating that no evident overfitting is observed.

To further test the transferability of the model to configurations outside the data distribution used in this work, the trained NEP model was used to predict atomic forces in the DP-potential training dataset constructed by Zong et al\cite{ZONG2026ijhmt}. Because this external dataset used pseudopotentials different from those used in the present work, the energy and virial values cannot be directly compared. Therefore, only the force prediction was evaluated. The force RMSE of the NEP model on this dataset is $41.5~\mathrm{meV/\text{\AA}}$, which is of the same order as the force errors obtained for the training and test datasets in this work. This result suggests that the NEP model provides additional evidence that the force prediction remains reasonable across a broader configuration space. Overall, the trained NEP model achieves good accuracy in predicting energy, force, and virial, and can serve as a reliable potential for the subsequent large-scale MD tensile simulations.

\begin{figure}
\centering
\includegraphics[width=1.0\columnwidth]{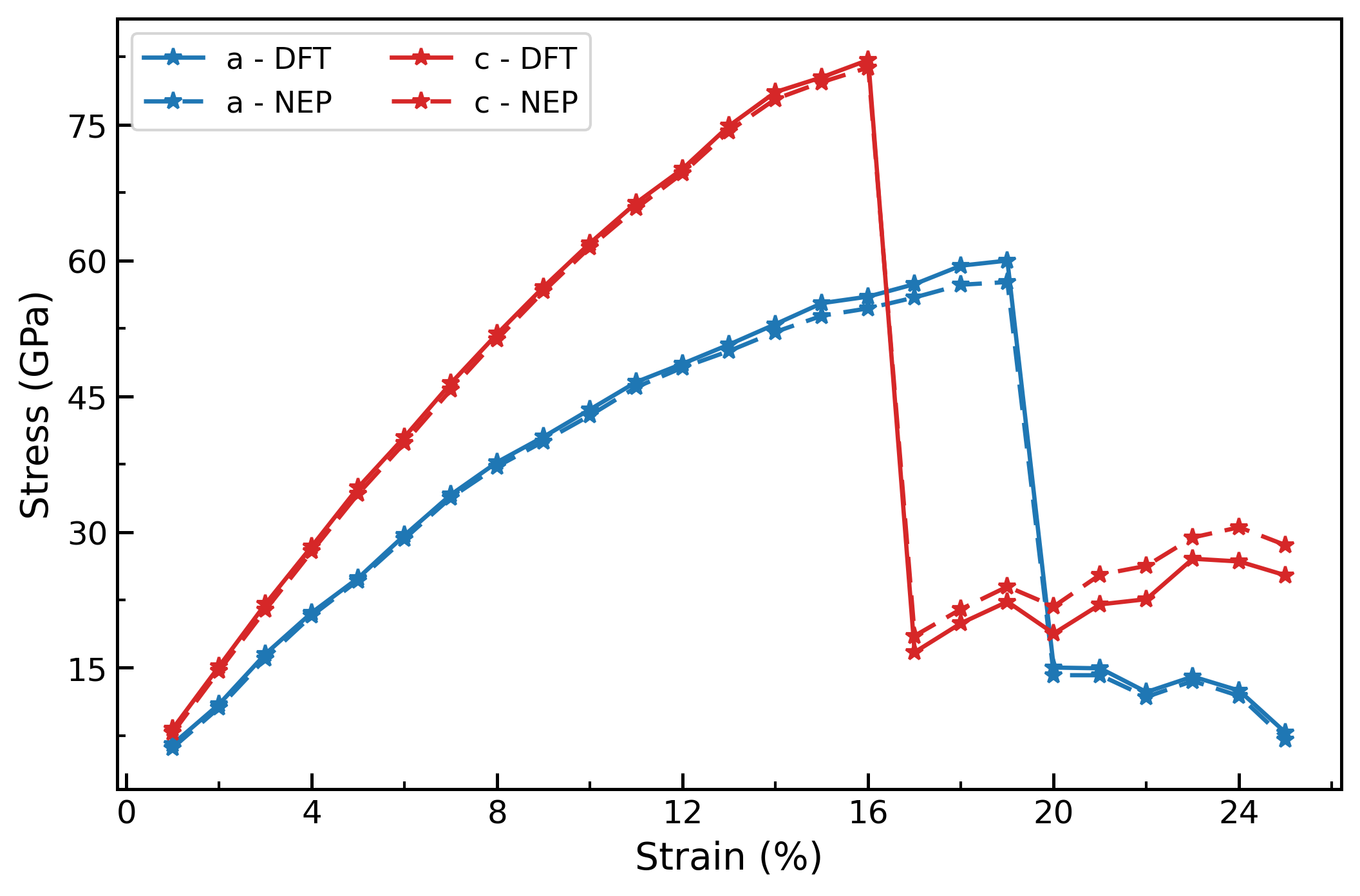}
\caption{\textbf{Comparison of NEP and DFT tensile responses of $\theta$-TaN.} Stress–strain curves for $a$-axis ($[2\bar{1}\bar{1}0]$) and $c$-axis ($[0001]$) tension at 300 K. The DFT data were obtained from static calculations on representative 128-atom configurations extracted from the NEP-MD tensile trajectories over the strain range of 1\%–25\%. }
\label{fig:nep2dft}
\end{figure}

To further verify the reliability of the NEP model under tensile loading, representative configurations were extracted from the 300 K NEP-MD tensile trajectories along increasing strain. Configurations from both the $a$-axis and $c$-axis tensile processes were recalculated by static DFT calculations and compared with the NEP predictions, as shown in Fig.~\ref{fig:nep2dft}. In the elastic regime, the stress--strain curves predicted by NEP almost overlap with the DFT reference data for both loading directions. No clear deviation is observed. Under $c$-axis tension, the material reaches a peak stress of approximately $82~\mathrm{GPa}$ at a strain of about $16\%$, followed by fracture and a sharp stress drop. Under $a$-axis tension, the fracture strain is higher, about $19\%$, whereas the peak stress is approximately $60~\mathrm{GPa}$, much lower than that along the $c$ axis. This confirms the pronounced mechanical anisotropy of $\theta$-TaN. The fracture strain and peak stress predicted by NEP are consistent with the DFT results for both directions. This agreement indicates that the NEP model can capture the tensile response of $\theta$-TaN up to fracture, as well as its anisotropic behavior. After fracture, the residual stresses predicted by the two methods show the same overall trend, with only small deviations. These deviations do not affect the description of the key tensile features, including the elastic response, fracture strain, and peak stress.

In addition to the static fitting accuracy and tensile response, we also examined the ability of the NEP model to describe key metastable configurations, including surface energies and generalized stacking fault energies. The results are summarized in Table~\ref{table:energy}. The surface energy was calculated as $\gamma_s = \frac{E_{\rm slab} - N E_{\rm bulk}}{2A},$ where $E_{\rm slab}$ is the total energy of the slab, $E_{\rm bulk}$ is the average atomic energy of the corresponding bulk structure, $N$ is the number of atoms in the slab, and $A$ is the surface area. For the three low-index surfaces, namely the basal plane $(0001)$, the $m$ plane $(10\bar{1}0)$, and the $a$ plane $(11\bar{2}0)$, the surface energies predicted by NEP agree well with the DFT values. The largest deviation is less than $0.14~\mathrm{J \cdot m^{-2}}$.

The generalized stacking fault energy (GSFE) was calculated by applying rigid shear displacements along selected slip planes to generate a series of faulted configurations. The maximum value of the GSFE curve corresponds to the unstable stacking fault energy, $\gamma_{\rm us}$, which is closely related to the difficulty of dislocation nucleation. In this work, $\gamma_{\rm us}$ was calculated for three slip systems: $(0001)[11\bar{2}0]$ basal slip, $(10\bar{1}0)[11\bar{2}0]$ $m$-plane slip, and $(11\bar{2}0)[0001]$ $a$-plane slip. The results are also listed in Table~\ref{table:energy}. For basal slip and $a$-plane slip, the NEP predictions agree well with the DFT reference values, with deviations of only $0.02~\mathrm{J \cdot m^{-2}}$ and $0.13~\mathrm{J \cdot m^{-2}}$, respectively. For $m$-plane slip, the predicted $\gamma_{\rm us}$ is $13.16~\mathrm{J \cdot m^{-2}}$, slightly higher than the DFT value of $11.84~\mathrm{J \cdot m^{-2}}$. The absolute deviation is about $1.3~\mathrm{J \cdot m^{-2}}$, corresponding to a relative deviation of approximately $11\%$. Overall, the NEP model shows good consistency with DFT in predicting both surface energies and stacking fault energies. This suggests that the model can also describe the energetics of non-equilibrium configurations such as surfaces and stacking faults.

\begin{figure*}
\centering
\includegraphics[width=1.8\columnwidth]{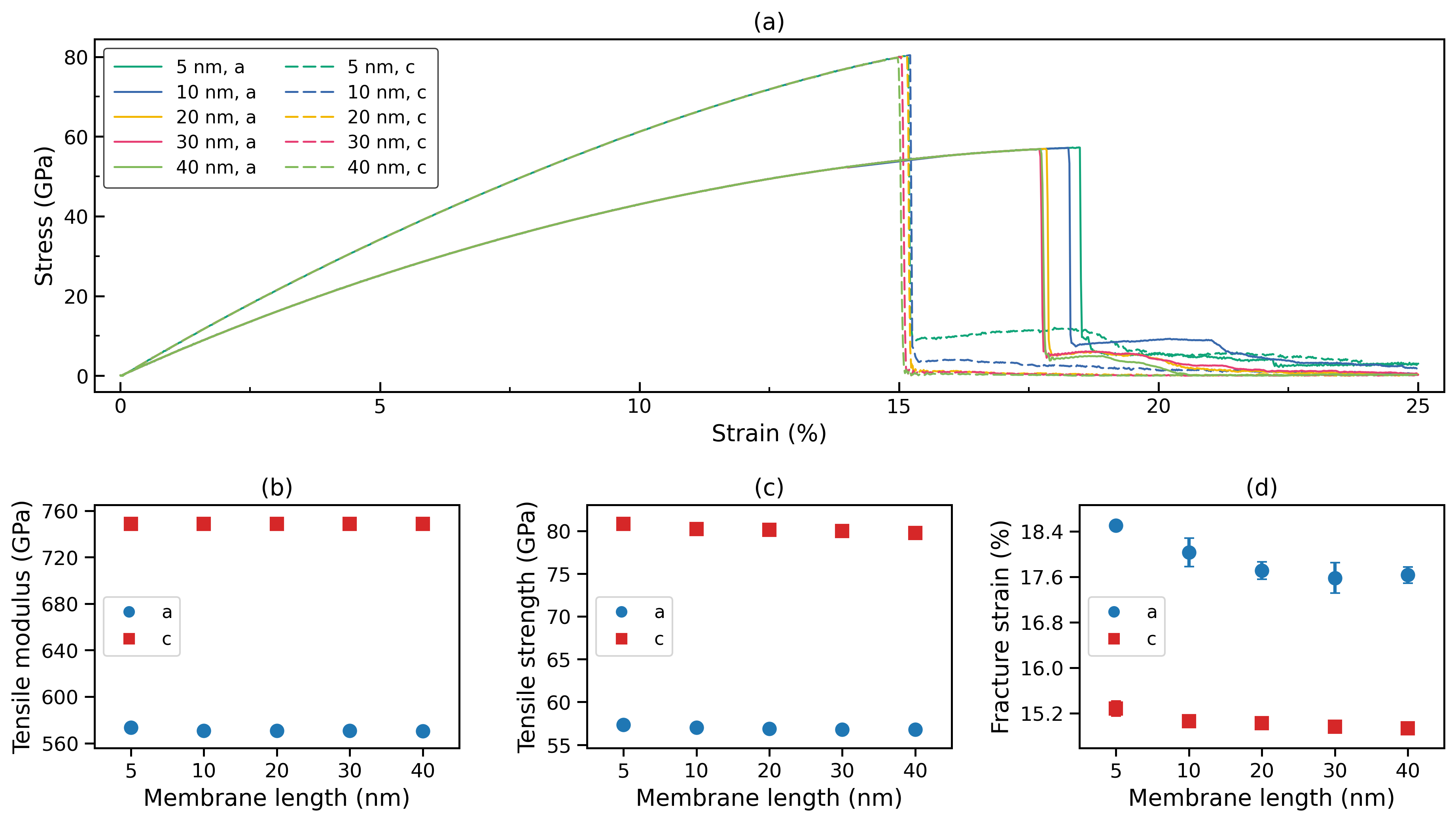}
\caption{\textbf{Size dependence of the tensile response of $\theta$-TaN.} (a) Stress–strain curves under tension along the $a$ axis ($[2\bar{1}\bar{1}0]$) and $c$ axis ($[0001]$) for different supercell sizes at 300 K and a strain rate of $1\times10^8~\mathrm{s^{-1}}$. (b–d) Tensile modulus, tensile strength, and fracture strain as functions of supercell size. Error bars represent the standard deviations from five independent simulations. The mechanical parameters converge when the supercell size reaches approximately $20~\mathrm{nm}$. }
\label{fig:size}
\end{figure*}

\begin{table}[htb]
\centering
\setlength{\tabcolsep}{1.5mm}
\caption{\textbf{Surface energies ($\gamma_{\mathrm{s}}$) and unstable stacking fault energies ($\gamma_{\mathrm{us}}$) of different crystallographic planes and slip directions from DFT and NEP.}}
\label{table:energy}
\begin{tabular}{cccccc}
\hline
\hline
Plane & \makecell{Slip \\ direction} & Type & \makecell{DFT \\ (J $\cdot$ m$^{-2}$)} & \makecell{NEP \\ (J $\cdot$ m$^{-2}$)} & Remark \\
\hline
$(0001)$       & ---            & $\gamma_{\mathrm{s}}$  & 4.18 & 4.14 & Basal \\
$(10\bar{1}0)$ & ---            & $\gamma_{\mathrm{s}}$  & 6.18 & 6.04 & $m$-plane \\
$(11\bar{2}0)$ & ---            & $\gamma_{\mathrm{s}}$  & 3.78 & 3.90 & $a$-plane \\
\hline
$(0001)$       & $[11\bar{2}0]$ & $\gamma_{\mathrm{us}}$ & 7.30 & 7.28 & Basal slip \\
$(10\bar{1}0)$ & $[11\bar{2}0]$ & $\gamma_{\mathrm{us}}$ & 11.84 & 13.16 & $m$-plane slip \\
$(11\bar{2}0)$ & $[0001]$       & $\gamma_{\mathrm{us}}$ & 6.49 & 6.36 & $a$-plane slip \\
\hline
\hline
\end{tabular}
\end{table}

\subsection{Size Convergence}
\label{size_sec}

\begin{table}[htb]
\centering
\setlength{\tabcolsep}{1.5mm}
\caption{\textbf{Simulation box dimensions and atom numbers of the $\theta$-TaN supercells used in the size-effect study.} $y$ and $z$ correspond to the loading directions along the $a$-axis ($[2\bar{1}\bar{1}0]$) and $c$-axis ($[0001]$), respectively.}
\label{table:size}
\begin{tabular}{cccccc}
\hline
\hline
Nominal size & $x$ (nm) & $y$ (nm) & $z$  (nm) & Atoms \\
\hline
5 nm  & 4.56  & 4.97  & 4.88  & 10,404 \\
10 nm & 9.63  & 9.95  & 9.77  & 87,856 \\
20 nm & 19.77 & 19.90 & 19.82 & 731,952 \\
30 nm & 29.90 & 29.85 & 29.88 & 2,503,488 \\
40 nm & 39.53 & 39.79 & 39.94 & 5,898,048 \\
\hline
\hline
\end{tabular}
\end{table}

To examine the effect of supercell size on the mechanical response of $\theta$-TaN, five supercells with nominal sizes of 5, 10, 20, 30, and $40~\mathrm{nm}$ were constructed. The actual box dimensions and the corresponding numbers of atoms are listed in Table~\ref{table:size}. For the supercells with sizes of $20~\mathrm{nm}$ and above, the actual box lengths in the three directions are nearly identical, which helps reduce the influence of box-shape anisotropy on the analysis of size effects. All tensile simulations were performed at $300~\mathrm{K}$ and a strain rate of $1\times10^8~\mathrm{s^{-1}}$ along both the $a$ and $c$ axes. For each size, five independent simulations were carried out to reduce the uncertainty caused by thermal fluctuations. The final results are reported as the average values and standard deviations of the five simulations.

Fig.~\ref{fig:size} shows the results of large-scale MD tensile simulations using the NEP potential. As shown in Fig.~\ref{fig:size}(a), the stress--strain curves for different supercell sizes almost overlap in the elastic regime and near the peak stress before fracture. The strength and stiffness along the $c$ axis are also much higher than those along the $a$ axis, consistent with the mechanical anisotropy observed in the DFT--NEP comparison. Fig.~\ref{fig:size}(b)--(d) show the tensile modulus, tensile strength, and fracture strain extracted from the stress--strain curves as functions of supercell size.

The tensile modulus and tensile strength show only weak size dependence. Under $a$-axis tension, the tensile modulus changes from $573.31\pm0.67~\mathrm{GPa}$ for the 5 nm model to $570.74\pm0.14~\mathrm{GPa}$ for the 20 nm model, while the tensile strength changes from $57.33\pm0.05~\mathrm{GPa}$ to $56.87\pm0.09~\mathrm{GPa}$. Under $c$-axis tension, the tensile modulus changes from $748.69\pm1.87~\mathrm{GPa}$ to $748.63\pm0.40~\mathrm{GPa}$, and the tensile strength changes from $80.79\pm0.28~\mathrm{GPa}$ to $80.10\pm0.23~\mathrm{GPa}$. These relative changes are all below $1\%$. In contrast, the fracture strain shows a more visible size dependence. Along the $a$ axis, it decreases from $18.50\pm0.07\%$ for the 5 nm model to $17.71\pm0.15\%$ for the 20 nm model. Along the $c$ axis, it decreases from $15.28\pm0.13\%$ to $15.02\pm0.09\%$. The relative changes in fracture strain are larger than those in modulus and strength.

\begin{figure*}
\centering
\includegraphics[width=1.8\columnwidth]{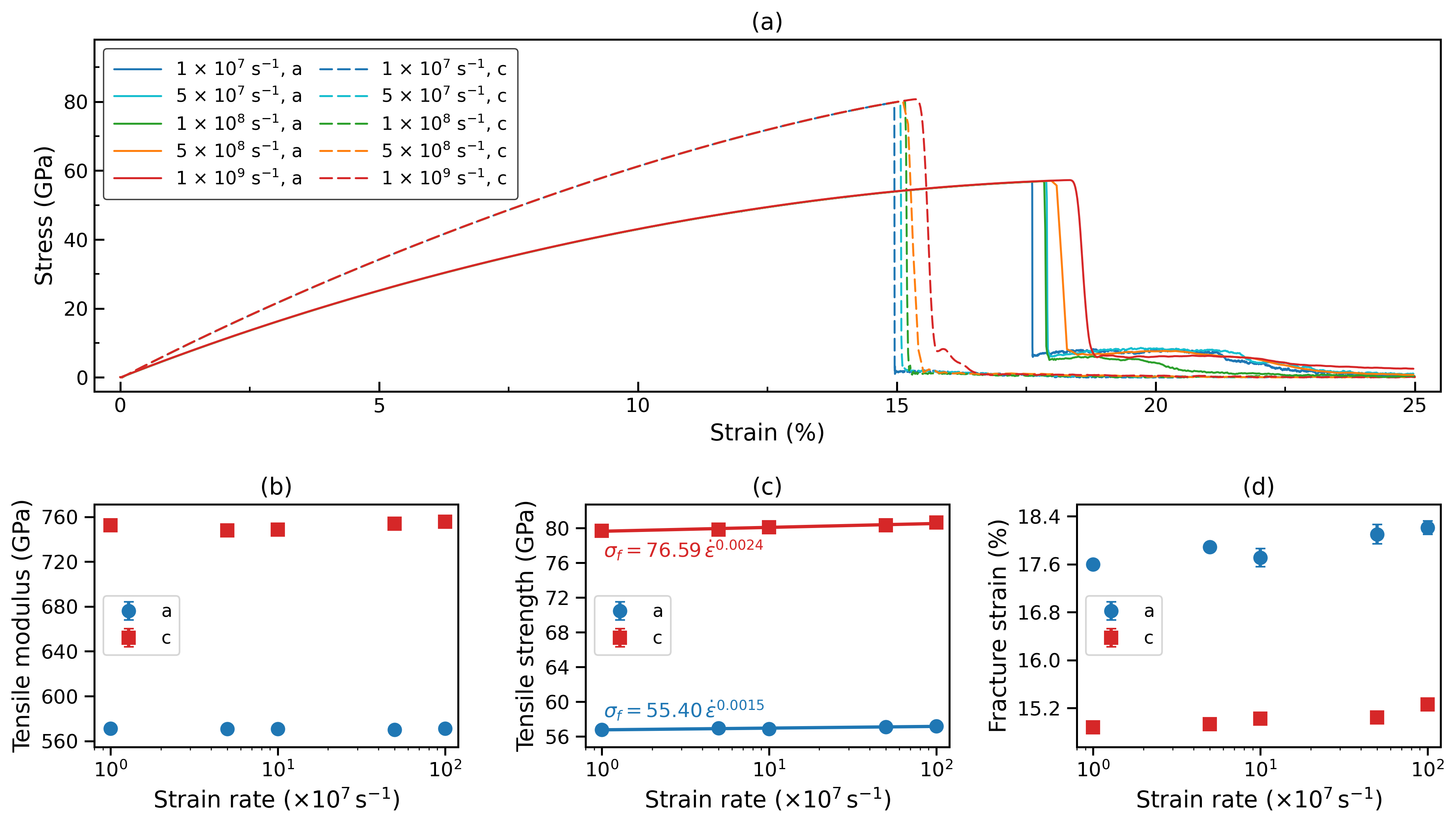}
\caption{\textbf{Strain-rate dependence of the tensile response of $\theta$-TaN.} (a) Stress–strain curves under tension along the $a$ axis ($[2\bar{1}\bar{1}0]$) and $c$ axis ($[0001]$) at different strain rates. The simulations were performed using the converged $20~\mathrm{nm}$ model at $300~\mathrm{K}$. (b–d) Tensile modulus, tensile strength, and fracture strain as functions of strain rate. Solid lines in (c) denote power-law fits of the tensile strength, $\sigma_f=C\dot{\varepsilon}^{m}$. Error bars represent the standard deviations from five independent simulations. }
\label{fig:speed}
\end{figure*}

Overall, all three mechanical parameters tend to converge when the supercell size reaches about $20~\mathrm{nm}$. Further increasing the size to 30 and $40~\mathrm{nm}$ does not lead to clear changes. This indicates that $20~\mathrm{nm}$ is a reasonable size threshold for reducing finite-size effects in tensile simulations of $\theta$-TaN. Therefore, the following analyses of temperature and strain-rate effects were carried out using this converged size. This trend is consistent with the size effects commonly observed in MD simulations of other materials\cite{YING2023EML}, where mechanical properties, especially fracture-related parameters, gradually converge with increasing simulation size. It also shows that sufficiently large supercells are needed to reduce finite-size effects in nanoscale mechanical simulations.

\subsection{Strain Rate Effect}
\label{defects}

After identifying $20~\mathrm{nm}$ as the converged size for reducing finite-size effects, we further examined the effect of strain rate on the tensile behavior of $\theta$-TaN along the $a$ and $c$ axes at $300~\mathrm{K}$. The strain rate ranged from $1\times10^7$ to $1\times10^9~\mathrm{s^{-1}}$, and the results are shown in Fig.~\ref{fig:speed}. 

Fig.~\ref{fig:speed}(a) presents the stress--strain curves at different strain rates. In the elastic regime, the curves almost overlap, indicating that the mechanical response in this stage is insensitive to strain rate. With increasing strain rate, the $a$-axis curves show a slight increase in peak stress and a small delay in fracture. In contrast, the $c$-axis curves are more closely grouped, suggesting an even weaker strain-rate effect. Fig.~\ref{fig:speed}(b) further shows that the tensile modulus remains nearly unchanged over the examined strain-rate range. The modulus is around $570~\mathrm{GPa}$ along the $a$ axis and between 750 and $756~\mathrm{GPa}$ along the $c$ axis, with no clear strain-rate dependence.

Fig.~\ref{fig:speed}(c) and (d) show the variation of tensile strength and fracture strain with strain rate, respectively. Similar to the tensile modulus, the tensile strength shows only a slight increase with increasing strain rate. Along the $a$ axis, it increases from $56.77\pm0.05~\mathrm{GPa}$ at $1\times10^7~\mathrm{s^{-1}}$ to $57.19\pm0.05~\mathrm{GPa}$ at $1\times10^9~\mathrm{s^{-1}}$. Along the $c$ axis, it increases from $79.72\pm0.12~\mathrm{GPa}$ to $80.64\pm0.19~\mathrm{GPa}$. The total increase is less than $1.5\%$ in both directions. This result indicates that the tensile strength of $\theta$-TaN is only weakly affected by strain rate in the range considered here. The fracture strain shows a more visible, but still moderate, strain-rate effect. With increasing strain rate, the fracture strain increases from $17.60\pm0.07\%$ to $18.21\pm0.11\%$ along the $a$ axis and from $14.88\pm0.05\%$ to $15.26\pm0.09\%$ along the $c$ axis. The corresponding increases are about $3.5\%$ and $2.6\%$, respectively.

The tensile strength data were further fitted using the power-law relation $\sigma_f = C\dot{\varepsilon}^{m}$ \cite{book1}. The fitted strain-rate sensitivity exponents are $m_a=0.0015$ for the $a$ axis and $m_c=0.0024$ for the $c$ axis, as shown in Fig.~\ref{fig:speed}(c). Both values are far below unity, further confirming that the tensile strength of $\theta$-TaN is insensitive to strain rate within the present range. The slightly larger value of $m_c$ indicates that the $c$-axis strength is marginally more strain-rate sensitive than the $a$-axis strength.

\subsection{Temperature Effect}
\begin{figure*}
\centering
\includegraphics[width=1.8\columnwidth]{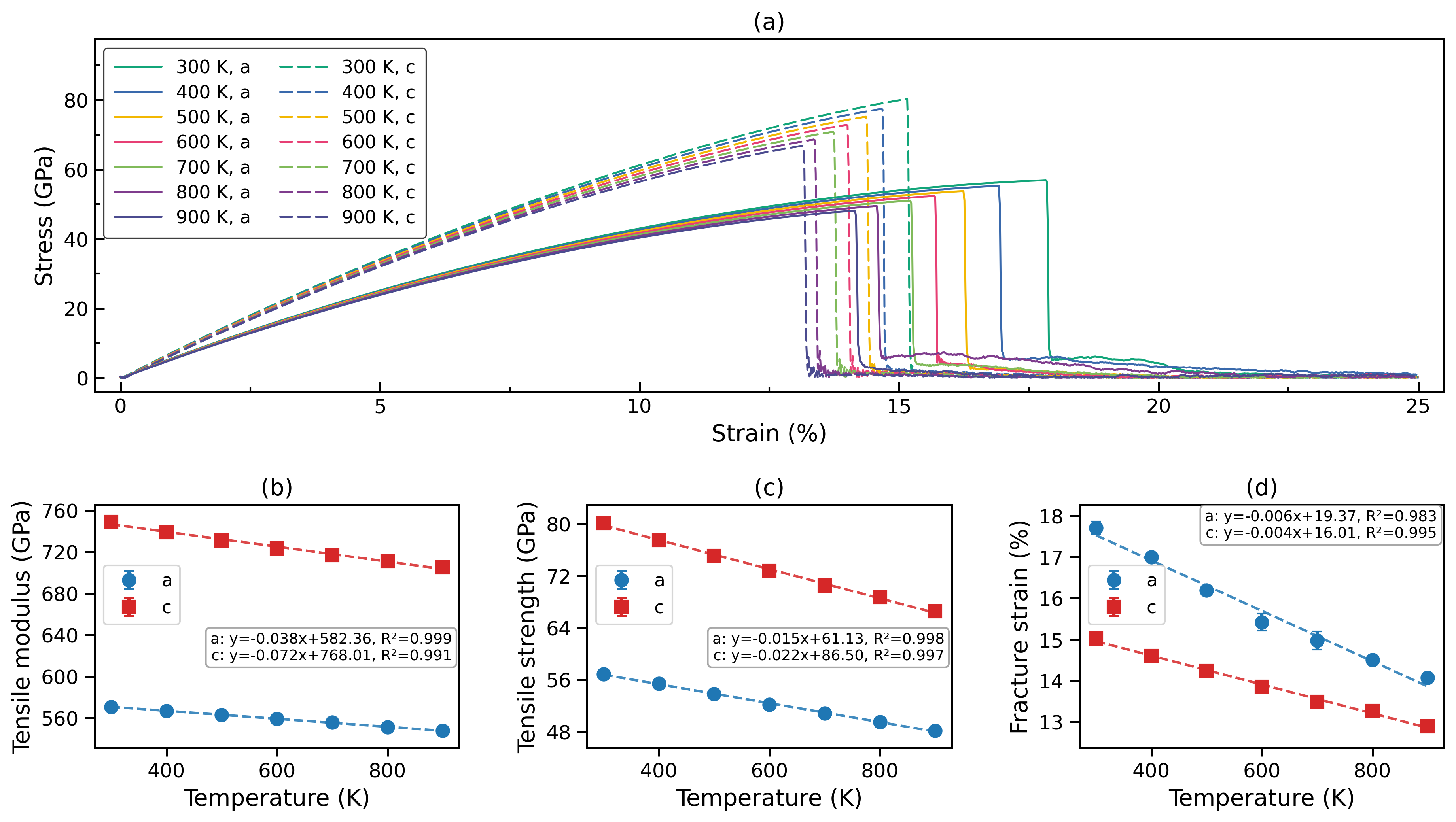}
\caption{\textbf{Temperature dependence of the tensile response of $\theta$-TaN.} (a) Stress–strain curves under tension along the $a$ axis ($[2\bar{1}\bar{1}0]$) and $c$ axis ($[0001]$) from 300 to $900~\mathrm{K}$. The simulations were performed using the converged $20~\mathrm{nm}$ model at a strain rate of $1\times10^8~\mathrm{s^{-1}}$. (b–d) Tensile modulus, tensile strength, and fracture strain as functions of temperature. Solid lines denote linear fits, and error bars represent the standard deviations from five independent simulations. }
\label{fig:tem}
\end{figure*}

Overall, in the strain-rate range of $1\times10^7$–$1\times10^9~\mathrm{s^{-1}}$, the tensile modulus and tensile strength of $\theta$-TaN show little dependence on strain rate, while the fracture strain increases only slightly with increasing strain rate. Compared with some two-dimensional materials, where both strength and fracture strain can increase strongly with strain rate, $\theta$-TaN shows a much weaker strain-rate dependence in this range. Its tensile response is therefore close to the strain-rate-insensitive limit.

To examine the effect of temperature on the tensile behavior of $\theta$-TaN, tensile simulations were performed from 300 to $900~\mathrm{K}$ at an interval of $100~\mathrm{K}$. The converged 20 nm model and a strain rate of $1\times10^8~\mathrm{s^{-1}}$ were used. Simulations were carried out along both the $a$ and $c$ axes. At each temperature, five independent runs were performed, and the averaged results are shown in Fig.~\ref{fig:tem}. 

As shown in Fig.~\ref{fig:tem}(a), increasing temperature reduces the slope of the stress--strain curves in the elastic regime for both directions. The peak stress and fracture strain also decrease, shifting the curves toward lower stress and lower strain. The $c$-axis curves remain above the $a$-axis curves over the whole temperature range, with no clear convergence or overlap between the two directions.

Figs.~\ref{fig:tem}(b)--(d) show the temperature dependence of the tensile modulus, tensile strength, and fracture strain. For the tensile modulus, the $a$-axis value decreases from $570.74\pm0.14~\mathrm{GPa}$ at 300 K to $547.71\pm0.45~\mathrm{GPa}$ at 900 K, corresponding to a decrease of about $4.0\%$. The $c$-axis value decreases from $748.63\pm0.40~\mathrm{GPa}$ to $705.09\pm0.42~\mathrm{GPa}$, corresponding to a decrease of about $5.8\%$. The decrease in tensile strength is more pronounced. Along the $a$ axis, the strength decreases from $56.87\pm0.09~\mathrm{GPa}$ at 300 K to $48.17\pm0.12~\mathrm{GPa}$ at 900 K, with a reduction of about $15.3\%$. Along the $c$ axis, it decreases from $80.10\pm0.23~\mathrm{GPa}$ to $66.51\pm0.30~\mathrm{GPa}$, with a reduction of about $17.0\%$. The fracture strain also decreases monotonically with temperature. It decreases from $17.71\pm0.15\%$ to $14.07\pm0.09\%$ along the $a$ axis, and from $15.02\pm0.09\%$ to $12.89\pm0.12\%$ along the $c$ axis. The corresponding reductions are about $20.6\%$ and $14.2\%$, respectively.

Linear fitting shows that the temperature dependences of the modulus, strength, and fracture strain can all be described by $y = kT + b$, with $R^2$ values higher than 0.98. This indicates that the tensile properties of $\theta$-TaN decrease almost linearly with increasing temperature in the range of 300–900 K.

The degradation of the mechanical properties can be attributed to thermal softening. Increasing temperature enhances atomic thermal vibrations and increases fluctuations around the equilibrium interatomic distances. This weakens the effective bonding strength of the Ta--N bonds to some extent. As a result, bond breaking can occur at lower applied stress and strain levels, leading to simultaneous decreases in tensile modulus, tensile strength, and fracture strain. This trend is consistent with the thermal softening behavior commonly observed in other material systems\cite{material2021acs,YING2023EML}.

\begin{figure*}
\centering
\includegraphics[width=1.8\columnwidth]{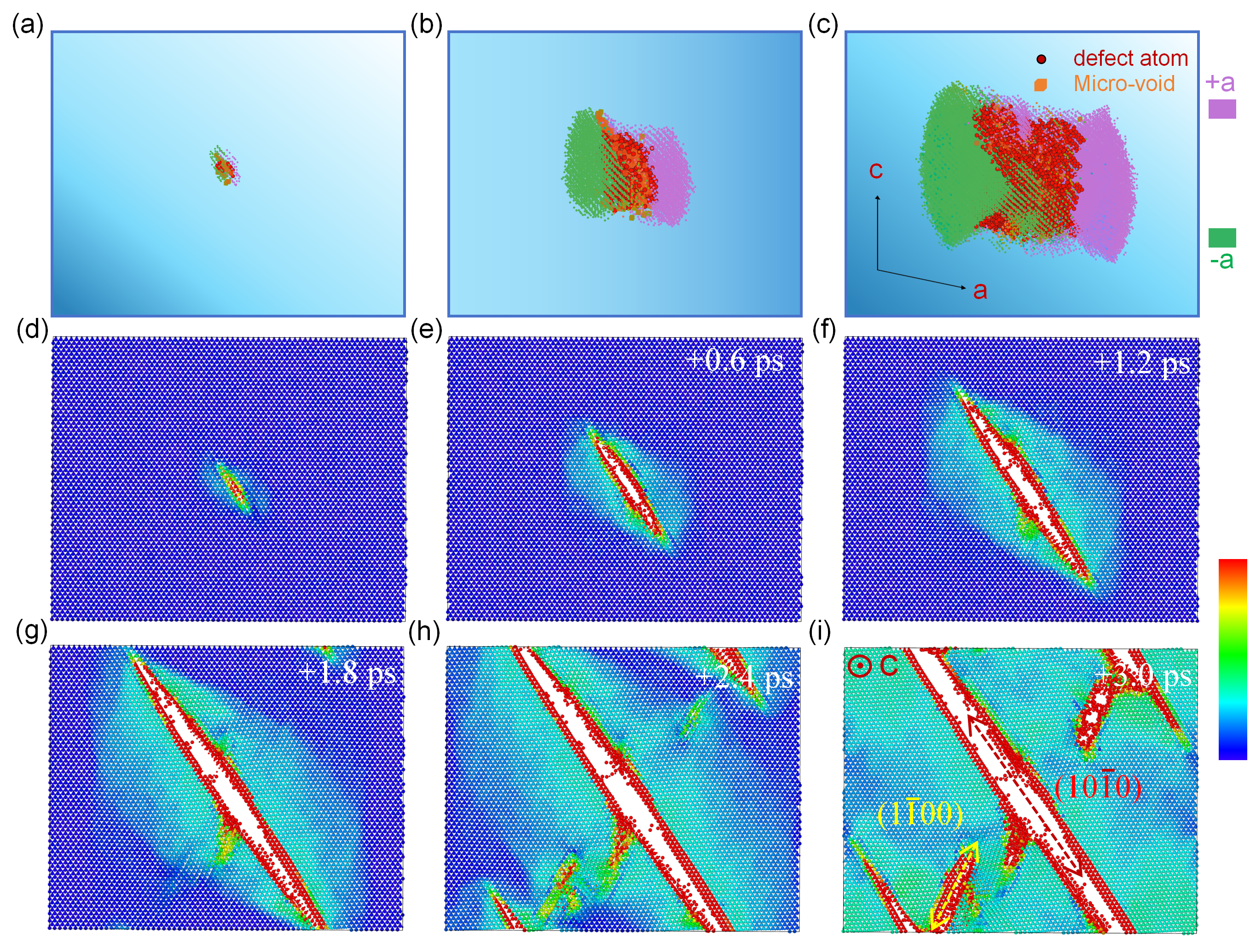}
\caption{\textbf{Atomic-scale fracture process of $\theta$-TaN under $a$-axis tension.} (a–c) Atomic displacement maps showing opposite displacement regions, defect atoms, and micro-voids near the crack-initiation region. Defect atoms are atoms with a normalized displacement larger than 0.5 and are used only for visualization. The micro-voids indicate local void formation after tensile deformation. (d–f) Sliced atomic configurations corresponding to the displacement maps in (a–c). (g–i) Sliced atomic configurations colored by local atomic shear strain, showing the subsequent crack-propagation process under tension along the $a$ axis ($[2\bar{1}\bar{1}0]$). The crack propagates along the $(10\bar{1}0)$ plane and then turns toward the equivalent $(1\bar{1}00)$ plane, both belonging to the ${10\bar{1}0}$ prismatic plane family. The time $t=0$ corresponds to the onset of visible crack formation. }
\label{fig:aaxes}
\end{figure*}

It should also be noted that the two loading directions do not show the same temperature sensitivity. For modulus and strength, the relative decreases along the $c$ axis are $5.8\%$ and $17.0\%$, slightly larger than the corresponding values of $4.0\%$ and $15.3\%$ along the $a$ axis. This suggests that the stiffness and strength along the $c$ axis are more sensitive to temperature. In contrast, the fracture strain shows the opposite trend. The reduction along the $a$ axis is $20.6\%$, larger than the $14.2\%$ reduction along the $c$ axis, indicating that the strain tolerance along the $a$ axis is more temperature sensitive. This difference between the temperature sensitivity of modulus/strength and that of fracture strain suggests that different local response mechanisms may operate along the two crystallographic directions during thermal softening. Nevertheless, over the whole temperature range of 300–900 K, the tensile modulus and tensile strength along the $c$ axis remain higher than those along the $a$ axis.

\subsection{Fracture mechanism and atomic displacement deviation}

\begin{figure*}
\centering
\includegraphics[width=1.8\columnwidth]{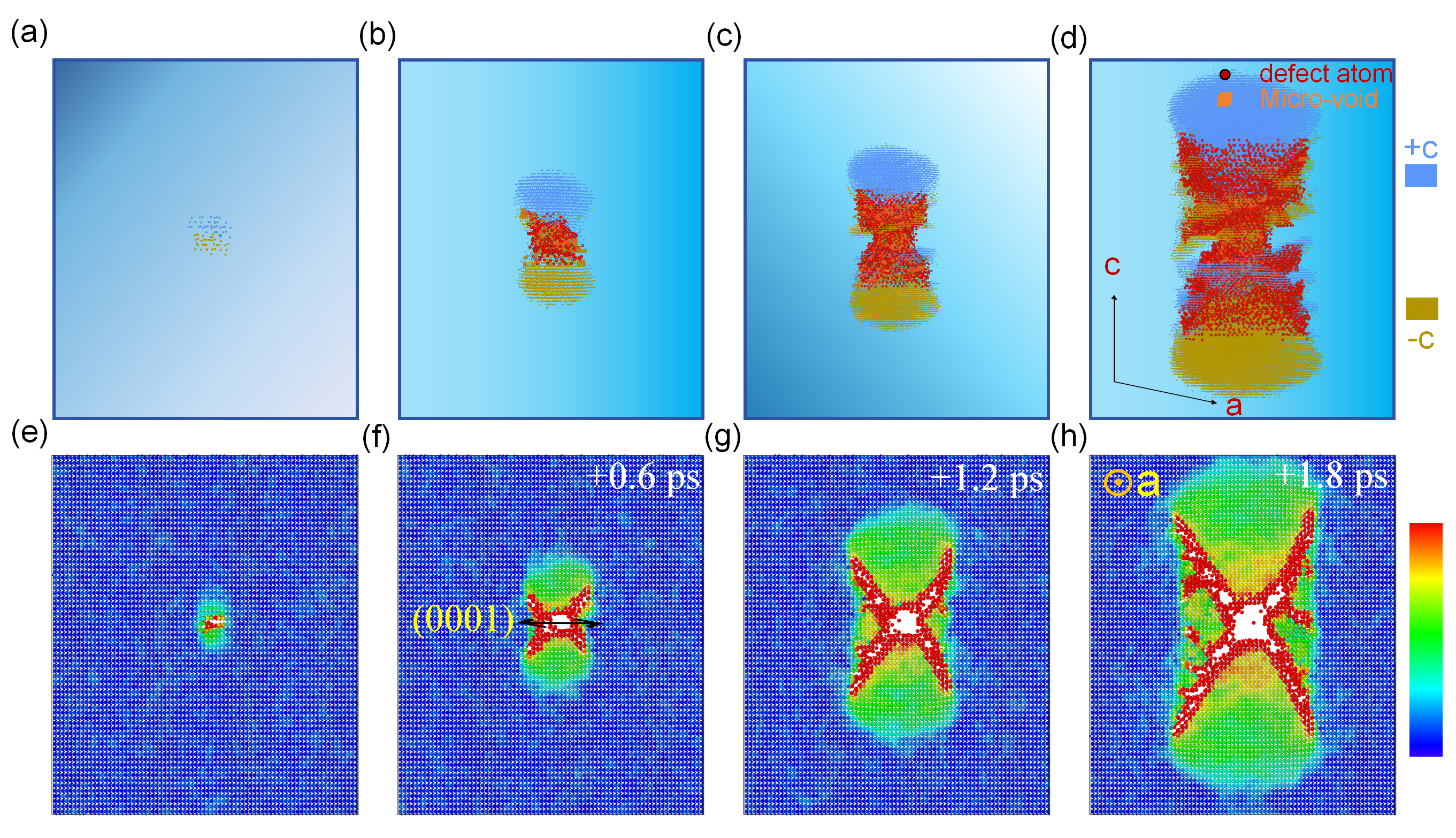}
\caption{\textbf{Atomic-scale fracture process of $\theta$-TaN under $c$-axis tension.}  (a–d) Atomic displacement maps showing localized displacement regions, defect atoms, and micro-voids near the crack-initiation region under tension along the $c$ axis ($[0001]$). Defect atoms are atoms with a normalized displacement larger than 0.5 and are used only for visualization. The micro-voids indicate local void formation after tensile deformation. (e–h) Sliced atomic configurations colored by local atomic shear strain, showing the subsequent crack-propagation process. Fracture initiates on the $(0001)$ basal plane and rapidly develops into a localized damage zone nearly symmetric about the loading axis. The time $t=0$ corresponds to the onset of visible crack formation. }
\label{fig:caxes}
\end{figure*}

The preceding sections show that the macroscopic mechanical response of $\theta$-TaN, including Young’s modulus, tensile strength, and fracture strain, changes with system size, strain rate, and temperature. They also show clear anisotropy under tensile loading along the $a$ and $c$ axes. However, these macroscopic parameters mainly describe the overall response of the material. They do not directly show how cracks nucleate and propagate at the atomic scale, nor do they explain the microscopic origin of the different fracture behaviors along the two loading directions. Therefore, the local structural evolution during tensile deformation needs to be examined.

During the tensile simulations, no evident dislocation slip or dislocation multiplication is observed in either loading direction. This indicates that the fracture of $\theta$-TaN is not controlled by dislocation-mediated plastic deformation. Together with the rapid stress drop after the peak stress and the absence of a clear plastic zone, this behavior suggests a brittle fracture mode. To examine this process in more detail, we analyzed the atomic displacement and local atomic shear strain during tensile loading.

The atomic displacement was calculated relative to the original equilibrium configuration. In the following analysis, atoms with a normalized displacement larger than 0.5 are labeled as “defect atoms” for visualization purposes. This term does not refer to crystallographic defects identified by a defect-recognition algorithm. Instead, it refers to atoms with relatively large displacement during tensile deformation. Therefore, the distribution of these defect atoms is used as a qualitative indicator of local displacement concentration and damage accumulation before crack initiation. The clustering of defect atoms, together with the formation of microvoids, marks the regions where local structural instability develops before macroscopic crack propagation. As a complementary measure, the local atomic shear strain field obtained from OVITO was used to describe lattice distortion around the propagating crack and its relation to the cleavage plane. Based on these analyses, the fracture processes under $a$-axis and $c$-axis tension are discussed separately below, followed by a comparison of their fracture paths.

We also tracked the average Ta--N bond length during tensile deformation. Before fracture, the average bond length changes only slightly and does not show clear progressive elongation. This is consistent with the strong bonding in $\theta$-TaN. Combined with the absence of evident dislocation activity, the rapid post-peak stress drop, and the lack of a clear plastic zone, this result suggests that failure in $\theta$-TaN does not occur through gradual weakening of the whole bonding network. Instead, it is more likely governed by sudden bond breaking triggered by local stress concentration.

Under $a$-axis tension, the atomic displacement maps in Fig.~\ref{fig:aaxes}(a)--(c) correspond to the sliced atomic configurations shown in Fig.~\ref{fig:aaxes}(d)--(f). These images show that atoms near the crack-nucleation region are displaced in opposite directions along the $a$ axis, forming the green and pink regions. This opposite-displacement pattern indicates that local displacement incompatibility has already developed before a visible crack forms. The defect atoms are mainly concentrated at the boundary between these two oppositely displaced regions, and microvoids also appear in the same region. This suggests that crack nucleation under $a$-axis tension is closely related to local displacement separation and void formation.

The sliced shear-strain maps in Fig.~\ref{fig:aaxes}(d)--(i) further show the subsequent crack-propagation process, where $t=0$ corresponds to the onset of crack formation. From $t=0$ to $3.0~\mathrm{ps}$, the high-shear-strain region expands with crack propagation and gradually localizes near the crack surfaces. At the same time, new defect clusters and microvoids continue to appear around the main crack. This indicates that damage accumulation proceeds together with crack growth, rather than being limited to the initial nucleation stage. In addition, the crack path shows clear crystallographic selectivity. The crack first propagates along the $(10\bar{1}0)$ plane and then turns toward the equivalent $(1\bar{1}00)$ plane. Both planes belong to the ${10\bar{1}0}$ prismatic plane family. These results show that, under $a$-axis tension, failure in $\theta$-TaN follows a sequence of local displacement separation, defect and microvoid accumulation, shear-strain localization, and crack propagation along selected cleavage planes.

Under $c$-axis tension, the atomic displacement maps in Fig.~\ref{fig:caxes}(a)--(d) show an evolution pattern different from that observed under $a$-axis tension. The displacement along the $c$ axis is concentrated in a relatively compact region that is nearly symmetric with respect to the loading axis, corresponding to the $+c$ and $-c$ displacement regions. Defect atoms and the associated microvoids are already concentrated in this localized region at an early stage, in contrast to the more dispersed distribution observed under $a$-axis tension.

The shear-strain maps in Fig.~\ref{fig:caxes}(e)--(h) show that fracture under $c$-axis tension first occurs on the $(0001)$ basal plane. After basal-plane cracking, the damaged region expands rapidly both along the $c$ axis and in the transverse direction, forming a spindle-shaped, cross-like damage zone. Defect atoms and microvoids spread from the initial crack region in multiple directions. The whole process occurs from $0$ to $1.8~\mathrm{ps}$, which is shorter than the $0$ to $3.0~\mathrm{ps}$ interval under $a$-axis tension. This indicates that fracture under $c$-axis loading is more abrupt, with damage that is more spatially localized. Unlike the inclined cleavage-plane propagation observed under $a$-axis tension, the damaged region under $c$-axis tension expands nearly symmetrically outward from the initial basal-plane crack and eventually forms a damage zone localized along the loading axis.

A comparison between the two loading directions shows that $\theta$-TaN has pronounced fracture anisotropy at the atomic scale. Under $a$-axis tension, atoms near the crack-nucleation region separate in opposite directions within the loading plane. Defect atoms and microvoids concentrate at the boundary between the two opposite-displacement regions. The crack then propagates along the crystallographically equivalent $(10\bar{1}0)$ and $(1\bar{1}00)$ planes, showing a relatively slow and inclined crack path controlled by selected cleavage planes. In contrast, under $c$-axis tension, fracture initiates directly on the $(0001)$ basal plane and rapidly develops into a spindle-shaped damage zone that is nearly symmetric about the loading axis. This corresponds to a faster and more spatially concentrated failure process.

These differences indicate that the fracture path of $\theta$-TaN is closely related to the cracking and propagation of specific crystallographic planes under different loading directions. Under $a$-axis tension, crack propagation mainly occurs along the ${10\bar{1}0}$ prismatic plane family. Under $c$-axis tension, crack initiation and damage growth mainly occur on and near the $(0001)$ basal plane. The relatively high unstable stacking fault energies indicate that dislocation nucleation is energetically unfavorable, which is consistent with the absence of obvious dislocation structures during tensile fracture. Combined with the evolution of local atomic displacement and shear strain, these results suggest that the anisotropic fracture behavior of $\theta$-TaN mainly arises from direction-dependent bond breaking and cleavage-plane selection, rather than from dislocation-mediated plastic deformation.

\subsection{Comparative strength analysis}

\begin{figure}
\centering
\includegraphics[width=1.0\columnwidth]{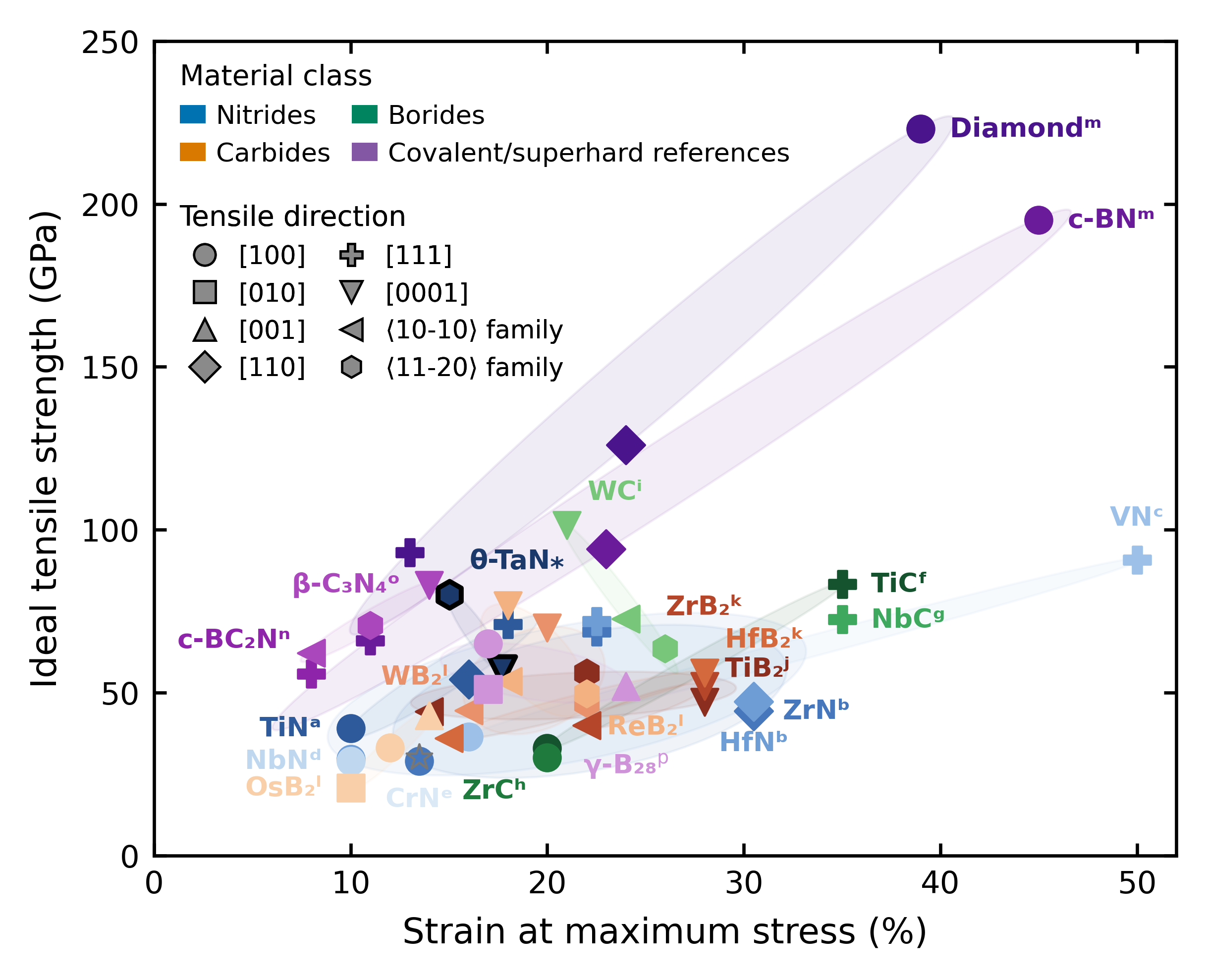}
\caption{\textbf{Comparison of ideal tensile strength and fracture strain for $\theta$-TaN and representative high-strength materials.} The ideal tensile strengths of $\theta$-TaN along the $a$ axis ($[2\bar{1}\bar{1}0]$) and $c$ axis ($[0001]$) are compared with reported values for representative nitrides, carbides, borides, WC, and covalent superhard materials. The corresponding fracture strain is shown for each material. Superscript letters denote the data sources: a, Ref.~\cite{refa}; b, Ref.~\cite{refb}; c, Ref.~\cite{refc}; d, Ref.~\cite{refd}; e, Ref.~\cite{refe}; f, Ref.~\cite{reff}; g, Ref.~\cite{refg}; h, Ref.~\cite{refh}; i, Ref.~\cite{refi}; j, Ref.~\cite{refj}; k, Ref.~\cite{refk}; l, Ref.~\cite{refl}; m, Ref.~\cite{refm}; n, Ref.~\cite{refn}; and o, Ref.~\cite{refo}; and p, Ref.~\cite{refp}.}
\label{fig:materials}
\end{figure}

To evaluate the relative mechanical performance of $\theta$-TaN among known superhard materials and transition-metal compounds, Fig.~\ref{fig:materials} summarizes the ideal tensile strengths and the corresponding strains calculated in this work. The results are compared with reported values for nitrides (TiN\cite{refa}, VN\cite{refc}, CrN\cite{refe}, ZrN\cite{refb}, HfN\cite{refb}, and NbN\cite{refd}), carbides (TiC\cite{reff}, ZrC\cite{refh}, and NbC\cite{refg}), borides (TiB$_2$\cite{refj}, ZrB$_2$\cite{refk}, HfB$_2$\cite{refk}, ReB$_2$\cite{refl}, WB$_2$\cite{refl}, and OsB$_2$\cite{refl}), tungsten carbide (WC)\cite{refi}, and representative covalent superhard materials, including diamond\cite{refm}, $c$-BN\cite{refm}, $\beta$-C$_3$N$_4$\cite{refo}, $c$-BC$_2$N\cite{refn}, and $\gamma$-B$_{28}$\cite{refp}.

$\theta$-TaN shows different strength--strain combinations along the two tensile directions. Under tension along $[2\bar{1}\bar{1}0]$, the ideal strength is $56.87~\mathrm{GPa}$ at a strain of 17.71\%. Along $[0001]$, the ideal strength increases to $80.10~\mathrm{GPa}$, while the corresponding strain decreases to 15.02\%. This indicates that the $c$-axis direction has a higher intrinsic strength but a lower strain tolerance. Compared with related nitrides, $\theta$-TaN is stronger than most transition-metal nitrides, such as TiN, CrN, NbN, ZrN, and HfN, most of which fall in the range of 30--70 GPa. VN reaches a strength level close to that of $\theta$-TaN only at a much larger strain of about 50\%. These comparisons suggest that $\theta$-TaN belongs to the high-strength group among nitrides and can reach high tensile stress at a moderate strain.

Compared with carbides, the $c$-axis strength of $\theta$-TaN is close to or higher than that of TiC, while its corresponding strain is lower. This means that $\theta$-TaN reaches a similar or higher strength at a smaller strain, reflecting its high intrinsic stiffness. The strengths of ZrC and NbC are lower than those of $\theta$-TaN along both tensile directions. Compared with borides, $\theta$-TaN also lies at a relatively high strength level. In particular, the $c$-axis strength of $80.10~\mathrm{GPa}$ is higher than those of several common borides, including HfB$_2$, TiB$_2$, and WB$_2$.

WC and $\theta$-TaN share the WC-type crystal structure, and therefore provide a useful isostructural comparison for separating the roles of crystal geometry and chemical bonding in mechanical response. When similar crystallographic directions are compared, WC shows a higher ideal strength and larger strain tolerance. Along $[0001]$, WC has an ideal strength of $101.3~\mathrm{GPa}$ at a strain of about 21\%, both higher than the values of $\theta$-TaN, $80.10~\mathrm{GPa}$ and 15.02\%. Along the $\langle 11\bar{2}0\rangle$ family, WC has a strength of $63.5~\mathrm{GPa}$ at a strain of about 26\%, again higher than the $56.87~\mathrm{GPa}$ and 17.71\% obtained for $\theta$-TaN. The reported result for WC along $[10\bar{1}0]$ gives a strength of $72.6~\mathrm{GPa}$ at a strain of about 24\%, leading to an anisotropic ranking of $[0001] > [10\bar{1}0] > \langle 11\bar{2}0\rangle$.

The comparison shows that WC and $\theta$-TaN have a similar anisotropic trend: both exhibit the highest strength along the $c$ axis ($[0001]$) and lower strength within the basal plane. This suggests that the “axially stronger and in-plane weaker” anisotropy may be related to the trigonal-prismatic coordination geometry in the WC-type structure, rather than being determined by one specific type of chemical bonding alone. However, WC has higher strength and strain tolerance than $\theta$-TaN along the corresponding directions. This indicates that, under the same structural motif and similar anisotropic pattern, the W--C bonding network may have a higher intrinsic strength limit and deformation tolerance than the Ta--N bonding network. Therefore, chemical bonding mainly affects the absolute level of mechanical properties, whereas crystal geometry may largely control the directional pattern of anisotropy. Although $\theta$-TaN is weaker than WC, it is still stronger than most cubic transition-metal nitrides, showing its relatively high mechanical performance within the transition-metal nitride family.

Compared with covalent superhard reference materials, the ideal strengths of $\theta$-TaN and WC are both lower than those of diamond, about $225~\mathrm{GPa}$, and $c$-BN, about $197~\mathrm{GPa}$. However, they are close to the strength range of metastable covalent superhard materials such as $\beta$-C$_3$N$_4$ and $c$-BC$_2$N. This indicates that these WC-type compounds do not reach the strength level of classical superhard materials, but they still fall within the range of high-ideal-strength materials. Overall, $\theta$-TaN occupies a position between conventional cubic transition-metal nitrides/carbides and covalent superhard materials in the strength--strain map. Its strength difference between $[2\bar{1}\bar{1}0]$ and $[0001]$ directions, $56.87~\mathrm{GPa}$ versus $80.10~\mathrm{GPa}$, together with the corresponding strain difference, 17.71\% versus 15.02\%, reflects pronounced mechanical anisotropy. The comparison with isostructural WC further shows that even when the crystal geometry is the same, chemical bonding can still play an important role in determining the ideal mechanical limits of a material.

\section{Conclusions}

In this work, we systematically investigated the tensile mechanical properties and fracture mechanism of the ultrahigh-thermal-conductivity metallic nitride $\theta$-TaN using machine-learning-potential molecular dynamics simulations. A reliable mechanical benchmark was established based on size convergence at $20~\mathrm{nm}$ and weak strain-rate sensitivity, with variations in all mechanical parameters below 3.5\% over the strain-rate range of $10^7$–$10^9~\mathrm{s^{-1}}$. The calculated results show that $\theta$-TaN exhibits pronounced tensile anisotropy. Along the $c$ axis ($[0001]$), it has a higher modulus of $748.63~\mathrm{GPa}$ and tensile strength of $80.10~\mathrm{GPa}$, but a lower fracture strain of 15.02\%, showing a response that is harder, stronger, and less strain tolerant. Along the $a$ axis ($[2\bar{1}\bar{1}0]$), the modulus and strength are lower, $570.74~\mathrm{GPa}$ and $56.87~\mathrm{GPa}$, respectively, whereas the fracture strain is higher, reaching 17.71\%. In the temperature range of 300–900 K, the modulus, strength, and fracture strain along both directions decrease almost linearly with increasing temperature. Even at $900~\mathrm{K}$, the material retains more than 73\% of its 300 K strength, without any abrupt transition or failure-related turning point. This relatively gradual thermal softening is not inconsistent with its ultrahigh thermal conductivity.

Atomic-scale analysis shows that no obvious dislocation structures were identified during fracture along either direction. Instead, failure is mainly affected by direction-dependent cleavage-plane selection. Under $a$-axis tension, cracks mainly propagate obliquely along the ${10\bar{1}0}$ prismatic plane family, leading to a relatively extended damage-accumulation process. Under $c$-axis tension, cracks initiate on the $(0001)$ basal plane and rapidly evolve into a nearly symmetric spindle-shaped damage zone, indicating a more abrupt fracture process. The statistical analysis further shows that local displacement separation and microvoid formation generally precede macroscopic crack growth. This suggests that the brittle fracture of $\theta$-TaN originates from local instability of the atomic structure or bonding network, rather than from homogeneous weakening of the whole structure.

Comparison with reported materials shows that the strength of $\theta$-TaN is higher than that of most related transition-metal nitrides and is close to or higher than that of some carbides and borides. It can therefore be regarded as a material with relatively high ideal strength. Compared with isostructural WC, $\theta$-TaN shows a similar anisotropic ranking, with higher strength along the axial direction and lower strength in the basal-plane direction. However, WC has higher strength and strain tolerance. This comparison suggests that the crystal geometry mainly controls the directional pattern of anisotropy, whereas the chemical bonding type, namely the Ta--N and W--C bonding networks, plays a key role in determining the absolute level of mechanical properties. These results provide an atomic-scale basis for evaluating the mechanical reliability of $\theta$-TaN as an emerging ultrahigh-thermal-conductivity material.

\vspace{0.5cm}

\noindent{\textbf{Data availability:}}
    Complete input and output files for the \gls{nep} training and testing are freely available at a Zenodo repository. 

\begin{acknowledgments}
Chenyang Cao gratefully acknowledges Zhicheng Zong for his fruitful discussions and valuable suggestions.
\end{acknowledgments}

\bibliography{Strain/refs_strain}

\end{document}